\documentclass[epj]{svjour}
\usepackage{latexsym}
\usepackage{graphics}
\usepackage[fleqn]{amsmath}
\usepackage{mathtools}
\usepackage{cuted}
\usepackage{color}

\begin{document}

\title{Observation of two $\mathcal{PT}$ transitions in an electric circuit with balanced gain and loss}
\author{Tishuo Wang\inst{1} \and Jianxiong Fang\inst{1}\and Zhongyi Xie\inst{1} \and Nenghao Dong\inst{1} \and Yogesh N Joglekar\inst{2} 
\thanks{e-mail: yojoglek@iupui.edu}
	\and Zixin Wang\inst{3}
\thanks{e-mail: wangzix@mail.sysu.edu.cn}
     \and Jiaming Li\inst{1}
    \thanks{e-mail: lijiam29@mail.sysu.edu.cn}%
	 \and Le Luo\inst{1}
\thanks{e-mail: luole5@mail.sysu.edu.cn}%
}                     

\institute{School of Physics and Astronomy, Sun Yat-Sen University,
	Zhuhai, Guangdong, China 519082\and Department of Physics, Indiana University Purdue University Indianapolis (IUPUI), Indianapolis, Indiana 46202, USA\and School of Eletronics and Information Technology, Sun Yat-Sen University,
	Guangzhou, Guangdong, China 510006}
\date{Received: date / Revised version: date}
%
\abstract{
We investigate $\mathcal{PT}$-symmetry breaking transitions in a dimer comprising two $LC$ oscillators, one with loss and the second with gain. The electric energy of this four-mode model oscillates between the two LC circuits, and between capacitive and inductive energy within each $LC$ circuit. Its dynamics are described by a non-Hermitian, $\mathcal{PT}$-symmetric Hamiltonian with three different phases separated by two exceptional points. We systematically measure the eigenfrequencies of energy dynamics across the three regions as a function of gain-loss strength. In addition to observe the well-studied $\mathcal{PT}$ transition for oscillations across the two $LC$  circuits, at higher gain-loss strength,  transition within each LC circuit is also observed. With their extraordinary tuning ability, $\mathcal{PT}$-symmetric electronics are ideally suited for classical simulations of non-Hermitian systems.
} 
\maketitle

\section{Introduction}
\label{intro}
Open classical systems with gain and loss, described by non-Hermitian Hamiltonians that are invariant under combined operations of parity and time reversal ($\mathcal{PT}$) have attracted considerable attention over the past two decades. This interest started with the seminal work by Carl Bender and co-workers~\cite{Bender98,Bender07}, and blossomed after their first realization in the optical domain~\cite{Guo09,Ruter10}. A $\mathcal{PT}$-symmetric Hamiltonian has a purely real spectrum when the non-Hermiticity is small and its eigenvectors are simultaneous eigenvectors of the antilinear $\mathcal{PT}$ operator with eigenvalue one. When the non-Hermiticy exceeds a finite threshold, the spectrum changes into complex conjugate pairs, and the $\mathcal{PT}$ operator maps an eigenstate into the eigenstate of its complex-conjugate eigenvalue~\cite{epjap13}. This transition, called $\mathcal{PT}$-symmetry breaking transition occurs at an exceptional point (EP), where both eigenvalues and corresponding eigenvectors of the non-Hermitian Hamiltonian coalesce. Over the past decade, systems with gain and loss, represented by positive and negative purely imaginary potentials, have been realized in a multitude of ``wave systems'' including electrical circuits~\cite{Schindler11,Chitsazi17,Montiel18,Chitsazi14,Assawaworrarit17,Chen18}, synthetic photonic lattices~\cite{Regensburger12}, and micro-ring resonators~\cite{Peng14,Xu16}. On the other hand, due to the quantum limits on noise in linear amplifiers~\cite{Caves82}, effective $\mathcal{PT}$-symmetric systems in the quantum regime have only been recently realized across multiple platforms such as ultracold atoms~\cite{Li19}, NV centers~\cite{Wu19},  superconducting qubit~\cite{Naghiloo19}, single photons~\cite{Xiao17,Bian19}, and atomic assembles~\cite{Zhang16}. 

Compared with the more challenging quantum realizations, classical setups with balanced gain and loss, or with mode-selective losses have many advantages. They are conceptually simpler, experimentally more accessible, and can have extraordinary tuning abilities that include interaction-induced nonlinearities, memory effects, and time delays. Thus, classical simulations of interesting quantum systems, especially systems with losses, gain, and noise~\cite{Franson16}, are of great interest. Among classical platforms, electronic circuits have experimentally advanced the studies of non-Hermitian physics with their low-cost configurations and enhanced sensitivity~\cite{Chen18,Xiao19}, novel devices and applications~\cite{Assawaworrarit17}, and the ability to simulate topological condensed matter phenomena~\cite{CHLee18}. In particular, the simplest $\mathcal{PT}$-symmetric electric dimer, i.e. one dissipative RLC oscillator coupled to a -RLC oscillator with gain, has been used to demonstrate many novel concepts in non-Hermitian, $\mathcal{PT}$-symmetric systems such as the $\mathcal{PT}$-symmetry breaking transition and scattering in static or time-periodic (Floquet) $\mathcal{PT}$-symmetric circuits~\cite{Schindler11,Chitsazi17,Montiel18,Lin12,Li18,Ramezani12}. 

\begin{figure}
	\centering
	\resizebox{0.45\textwidth}{!}{%
		\includegraphics{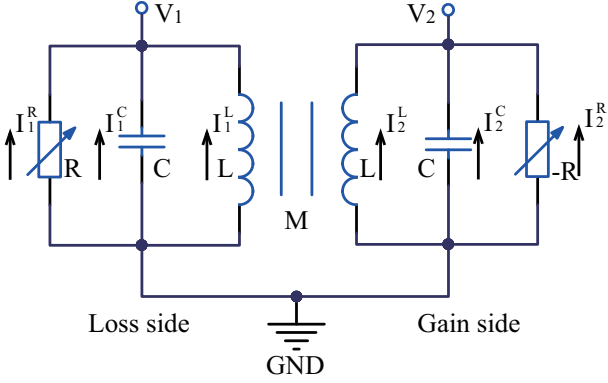}
	}
	\caption{Schematic of a $\mathcal{PT}$-symmetric electric circuit. The two $LC$ oscillators have positive and negative resistors $\pm R$, giving loss and gain respectively, and are coupled by the mutual inductance $M\geq 0$. The four-mode energy dynamics in this circuit is described by a $\mathcal{PT}$-symmetric, non-Hermitian Hamiltonian with two exceptional point degeneracies.}
	\label{Fig1}       
\end{figure}

Fig.~\ref{Fig1} illustrates a $\mathcal{PT}$-symmetric dimer~\cite{Schindler11}. It consists of two $LC$ oscillators, one with resistance $R$ (loss) and the other with effective, negative resistance $-R$ (gain), that are coupled via their mutual inductance $M$. Within an isolated $LC$ oscillator, the energy switches from the capacitor $C$ to the inductor $L$ with twice of frequency $\omega_0=1/\sqrt{LC}$. Due to the mutual inductance, the energy also oscillates between the gain-$LC$ circuit and the loss-$LC$ circuit with a much smaller frequency $\omega_D\sim\mu\omega_0<\omega_0$ where $0\leq\mu=M/L< 1$ denotes the dimensionless inductive coupling. Thus, we note that although Fig.~\ref{Fig1} presents the simplest electrical circuit with $\mathcal{PT}$-symmetry, it is not a prototypical two-mode $\mathcal{PT}$-symmetric system~\cite{Ruter10,Regensburger12,Peng14} with a single frequency scale; instead, it is a four-mode system with dynamics that are determined by the two frequency scales $\omega_D$ and $\omega_0$ respectively. As the dimensionless gain-loss strength of the circuit, given by $\gamma=\omega_0\tau_{LR}=\sqrt{L/C}/R$, is increased, the system transitions through two second-order EP degeneracies~\cite{Schindler11,Montiel18} instead of the single second-order EP transition seen in a two-mode system. Here $\tau_{LR}=L/R$ is the characteristic decay-time of an $LR$ circuit. The first of these transitions, occurring near $\gamma\sim\mu$, has been extensively studied and experimentally observed in the literature~\cite{Schindler11}. The second transition occurs near at much larger gain-loss strength, i.e $\gamma\sim 2$, and has not been experimentally observed. 

Here we report an experimental determination of the complete $\mathcal{PT}$-phase diagram of $\mathcal{PT}$-symmetric electric circuit dimer.  The plan of the paper is as follows. In Sec.~\ref{sec:theory} we present a physically transparent approach that allows us to translate Kirchoff-law dynamical equations for the $\mathcal{PT}$-dimer into a Schrödinger-like equation for dynamics of energy stored in the two capacitors and inductors. With the non-Hermitian, $\mathcal{PT}$-symmetric Hamiltonian for the $\mathcal{PT}$ electrical circuit, we obtain the well-known results~\cite{Schindler11,Montiel18} for $\mathcal{PT}$ transition thresholds. Experimental results for the eigenfrequencies, including the observation of the second EP degeneracy are discussed in Sec.~\ref{sec:expt}. This observation, at high gain-loss strength, requires a delicate control of the resistance balance, and its details are also presented in Sec.~\ref{sec:expt}. We conclude the paper with a brief discussion in Sec.~\ref{sec:disc}. 

\section{Theoretical analysis}
\label{sec:theory}

The dynamics of currents and voltages in Fig.~\ref{Fig1} are determined by Kirchoff laws~\cite{Schindler11}, 
\begin{eqnarray}
\label{eq: Kirch1}
0&=&I_{n}^{R}+I_{n}^{C}+I_{n}^{L},\\
V_{n}&=&(-1)^{n}I_{n}^{R}R=-\frac{1}{C}\int_{0}^{t}I_{n}^{C}(\tau)d\tau,\\
-V_{1}&=& L\frac{dI_{1}^{L}}{dt}+M\frac{dI_{2}^{L}}{dt},\\
\label{eq: Kirch4}
-V_{2}&=&L\frac{dI_{2}^{L}}{dt}+M\frac{dI_{1}^{L}}{dt},
\end{eqnarray}
where $n=1 (2)$ represents the loss (gain) oscillator, and the currents across the resistor, capacitor, and inductor are denoted by $I_{n}^{R}$, $I_{n}^{C}$ and $I_{n}^{L}$ respectively. Eqs.(\ref{eq: Kirch1})-(\ref{eq: Kirch4}) can be rewritten into a set of four, linear, coupled differential equations, 
\begin{eqnarray}
\label{eq: differ1}
\frac{dV_1}{dt}&=&-\gamma\omega_{0}V_1+\frac{I_{1}^L}{C},\\
\frac{dV_2}{dt}&=&\gamma\omega_{0}V_2+\frac{I_{2}^L}{C},\\
\frac{dI_1^L}{dt}&=&\frac{-V_1+\mu V_2}{L(1-\mu^2)},\\
\label{eq: differ4}
\frac{dI_{2}^L}{dt}&=&\frac{\mu V_1-V_2}{L(1-\mu^2)}.
\end{eqnarray}
Eqs.(\ref{eq: differ1})-(\ref{eq: differ4}) can be rewritten into a Schrodinger-like equation for the column vector $|\psi(t)\rangle=(V_1,V_2,I_{1}^L,I_{2}^L)^T$, i.e. $id|\psi\rangle/dt=h|\psi\rangle$ where $h$ is a $4\times4$ non-Hermitian matrix with purely imaginary entries~\cite{Schindler11}, 
\begin{eqnarray}
\label{eq: Shrodinger}
h& =& i\left[
\begin{array}{cccc} 
-\gamma\omega_0&0&\frac{1}{C}&0 \\ 
0&\gamma\omega_0&0&\frac{1}{C} \\ 
-\frac{1}{L(1-\mu^2)}&  \frac{\mu}{L(1-\mu^2)} & 0 & 0 \\ 
\frac{\mu}{L(1-\mu^2)}&-\frac{1}{L(1-\mu^2)} & 0 & 0
\end{array}
\right]. 
\end{eqnarray}
We note that casting the Kirchoff-law equations into a Schrodinger-like form is neither physically transparent not very useful, because the matrix $h$ does not become Hermitian even in the no-gain, no-loss limit ($\gamma=0$). Moreover, since the vector $|\psi\rangle$ has entries with two different engineering dimensions, so does the matrix $h$. To make a connection with the $\mathcal{PT}$-symmetric quantum theory, we need to identify a quantity that remains conserved in the $\gamma=0$ limit and study its dynamics. For two inductively coupled $LC$ oscillators, that quantity is given by the circuit energy, i.e $\mathcal{Q}(t)=\langle\psi(t)|A|\psi(t)\rangle$  where the positive-definite, bilinear form $A$ is given by 
\begin{eqnarray}
\label{eq: energy}
A&=&\left( 
\begin{array}{cccc} 
\frac{C}{2}&0&0&0 \\ 
0&\frac{C}{2}&0&0 \\ 
0&0&\frac{L}{2}&\frac{1}{2}\mu L \\ 
0&0&\frac{1}{2}\mu L&\frac{L}{2} 
\end{array}
\right).
\end{eqnarray}
Thus, an analogy with the standard quantum theory (where the norm a state is conserved in the Hermitian limit), we consider a state whose norm is given by the energy, i.e. $|\phi\rangle=A^{1/2}|\psi\rangle$ where $A^{1/2}$ denotes the positive, Hermitian square root of the matrix $A$. We also note that all elements of the column vector $|\phi\rangle$ now have the same engineering dimensions. With this change of basis, the Kirchoff-law equations can be written as 
\begin{eqnarray}
\label{eq: se1}
i\frac{d|\phi\rangle}{dt} &=&H|\phi\rangle ,\\
\label{eq: se2}
H(\gamma)&=&\frac{\omega_{0}}{2}\left( 
\begin{array}{cccc} 
-2i\gamma&0&i\gamma_{0}&-i\gamma_{\mathcal{PT}} \\ 
0&2i\gamma&-i\gamma_{\mathcal{PT}} & i\gamma_{0} \\ 
-i\gamma_{0}&i\gamma_{\mathcal{PT}} &0  &0 \\ 
i\gamma_{\mathcal{PT}}&-i\gamma_{0}  &0  &0 
\end{array}
\right),
\end{eqnarray}
where $\gamma_{\mathcal{PT}}=1/\sqrt{1-\mu}-1/\sqrt{1+\mu}$ and $\gamma_{0}=1/\sqrt{1-\mu}+1/\sqrt{1+\mu}$ denote the locations of the two EPs, as we will discuss in the following paragraphs. Thus, the time-evolution of the ``square-root of energy'' state vector $|\phi\rangle$ is given by the Hamiltonian $H(\gamma)=A^{1/2}hA^{-1/2}$ that becomes Hermitian in the limit $\gamma=0$, and satisfies $[\mathcal{PT},H]=0$. Here the parity operator exchanges the labels $1\leftrightarrow 2$, and the time-reversal operator, in addition to complex conjugation ($\ast$), reverses the currents $I^L_n\rightarrow-I^L_n$. In matrix representation, they are given by 
\begin{eqnarray}
\label{eq: PT oper}
\mathcal{P}=\left( 
\begin{array}{cccc} 
0&1&0&0 \\ 
1&0&0&0 \\ 
0&0&0&1 \\ 
0&0&1&0 
\end{array}
\right), && 
\mathcal{T}=\left( 
\begin{array}{cccc} 
1&0&0&0 \\ 
0&1&0&0 \\ 
0&0&-1&0 \\ 
0&0&0&-1 
\end{array}
\right)\ast.
\end{eqnarray}

The eigenvalues of the $\mathcal{PT}$-symmetric Hamiltonian $H(\gamma)$, Eq.(\ref{eq: se2}), determine whether the system is in the $\mathcal{PT}$- symmetric phase (all real eigenvalues) or broken phase (emergence of complex conjugate eigenvalues), and in the latter case, the degree of $\mathcal{PT}$-symmetry breaking i.e. the fraction of eigenvalues that have become complex~\cite{Joglekar10,Scott11,Jake11}. They are given by 
\begin{eqnarray}
\label{eq: eigenfreq1}
\omega_{1,2}& = &\frac{\omega_0}{2}\left( \sqrt{\gamma_0^2-\gamma^2}\pm\sqrt{\gamma_{\mathcal{PT}}^2-\gamma^2}\right), \\
\omega_{3,4} & = & -\omega_{1,2}.
\end{eqnarray}
In the Hermitian limit $\gamma\rightarrow 0$, the eigenvalues reduce to $\omega_1=\omega_0/\sqrt{1-\mu}\geq \omega_2=\omega_0/\sqrt{1+\mu}$ as is expected for two coupled $LC$ oscillators. They are purely real when $\gamma\leq\gamma_\mathcal{PT}$. As $\gamma$ is increased from zero, the frequencies $\omega_{1,2}$ approach each other, and become degenerate at the first EP, $\gamma_\mathcal{PT}$. Then they turn into complex-conjugate pairs, as do $\omega_{3,4}=-\omega_{1,2}$. As $\gamma$ is increased further, the amplifying mode frequencies $\omega_{1,4}$ become degenerate at the second EP, $\gamma=\gamma_0$, as do the decaying-mode frequencies $\omega_{2,3}$. We note that for weakly coupled $LC$ oscillators, $\mu=M/L\ll 1$, a Taylor expansion gives $\omega_{1,2}\approx\omega_0(1\pm\mu/2)$, a Hermitian gap $\omega_1-\omega_2=\omega_0\mu$ and a dimensionless $\gamma_\mathcal{PT}\sim\mu$ thus recovering the $\mathcal{PT}$-threshold result for a prototypical dimer Hamiltonian $H_D=\omega_0(\mu\sigma_x+i\gamma\sigma_z)/2$. For $\mu\ll1$, the second threshold gives $\gamma_0\approx 2$ or equivalently $R_0=\sqrt{L/C}/2$. This is precisely the resistance at which a parallel $RLC$ circuit is critically damped. In the opposite limit, $\mu\rightarrow 1$, the two $LC$ circuits are strongly coupled, leading to a divergent $\gamma_\mathcal{PT}\approx \sqrt{L/\Delta L}$ and $\gamma_0=\gamma_\mathcal{PT}+\sqrt{2}$, where $\Delta L=L-M=L(1-\mu)$ denotes the small difference between the self-inductance and the mutual inductance. In this limit, the frequencies $\omega_{1,2}$ are asymmetrically located around $\omega_0$, with their difference diverging as $\omega_1-\omega_2=\omega_0\sqrt{L/\Delta L}$. 

In the parameter range $\gamma_\mathcal{PT}<\gamma<\gamma_0$, the energy dynamics across the two $LC$ circuits is in the $\mathcal{PT}$-broken phase while the dynamics within each $LC$ circuit is in the $\mathcal{PT}$-symmetric phase. Past the second EP at $\gamma=\gamma_0$, all four eigenvalues are purely imaginary, which indicates overdamped (dissipative) modes for the $RLC$ circuit and their gain analog for the $-RLC$ circuit. In this regime, dynamics across the two $LC$ circuits, as well as within each $LC$ circuit are in the $\mathcal{PT}$ broken phase. In our experiments, discussed in the following section, we only focus on the positive real and imaginary parts of the eigenvalues, as the remaining two eigenvalues are determined by the particle-hole symmetric nature of the spectrum~\cite{phsymmetry10}. 

\section{Experimental setup and results}
\label{sec:expt}

\begin{figure*}[t!]
	\centering
	\resizebox{1.0\textwidth}{!}{%
		\includegraphics{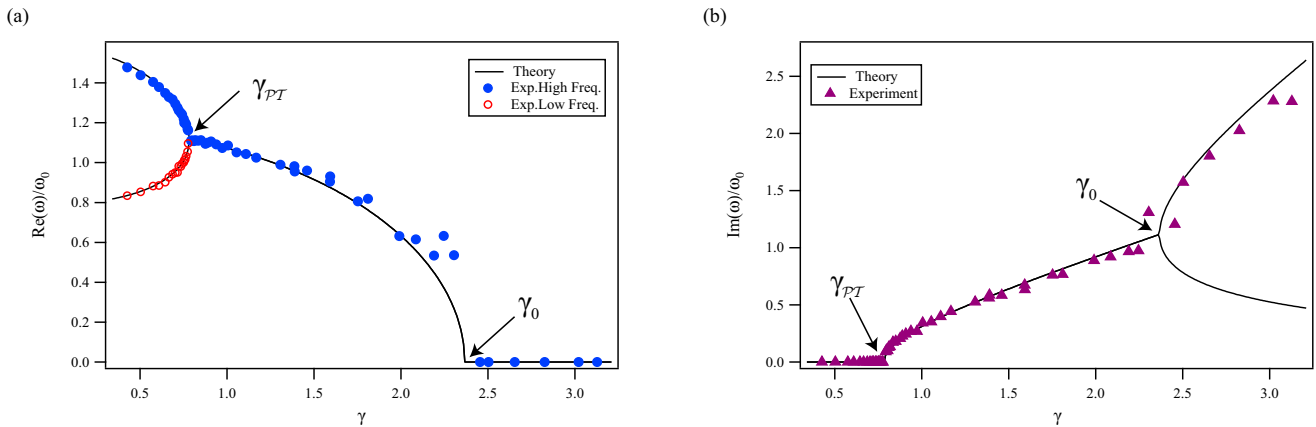}
	}
	\caption{Measurement of the complex eigenfrequencies $\omega_{1,2}(\gamma)/\omega_0$ as a function of dimensionless gain-loss $\gamma$ across the two exceptional points (EPs) occuring at $\gamma_\mathcal{PT}$ and $\gamma_0$ respectively. Symbols denote experimental results; theoretical predictions are shown by solid black lines. Although the transition at $\gamma_\mathcal{PT}$ is well-studied~\cite{Schindler11,Schindler13}, the transition across the second EP at $\gamma_0$ has not been observed in the past.
}
	\label{Phase1}       
\end{figure*}

The experimental setup is similar to that in Refs.~\cite{Schindler11,Schindler13}, although in those works, only the transition across the first EP at $\gamma_{\mathcal{PT}}$ was studied. However, the circuit in Fig.~\ref{Fig1} inherently allows exploration of all three regions, separated by the two EPs. Our $LC$ circuits have $L=7.91$ mH and $C=10.14$ nF, and a fundamental frequency of $\omega_0=2\pi\times 17.77$ kHz. The two oscillators are moderately coupled, i.e. $\mu=0.6$, leading to $\gamma_\mathcal{PT}=0.79$ and $\gamma_0=2.37$. To measure the eigenfrequencies across these regions, we adopt two methods. 

First, the inherent, resistive losses of the two inductors on the gain and the loss sides are canceled by two negative resistors. Without this cancellation, the experimental results do not match the theoretical analysis presented in the preceding section. In our setup, this resistance in each inductor, shown in Fig.~\ref{Fig1}, is $R_L=16.80 \rm\Omega$. It is compensated, in each isolated $\pm RLC$ circuit, by a negative impedance converter (NIC), which is based on a non-inverting amplifier~\cite{Schindler11,Schindler13}. When each isolated $LC$ circuit reaches this balance, the output voltage in each, i.e. $V_{1,2}(t)$ start self-oscillations~\cite{Schindler13}. In our experiments, the negative resistance compensates the inherent resistance of the inductors to within 2\% accuracy. We emphasize that although $R_L$ is tens of Ohms, its effects become important at large $\gamma$, which is achieved by reducing the loss-gain resistance $\pm R$. For example, to traverse across the second EP at $\gamma_0$, the loss-gain resistances $\pm R$ are reduced to 350 $\rm\Omega$. With an uncompensated inductor resistance, according to our simulations and experiments, this leads to a a 5\% shift in the location of $\gamma_0$. Thus, in order to accurately map out the location of the second EP, the inherent resistances in gain and loss inductors both need to be compensated. We note that the $\sim$kHz natural frequency of our setup reduces the effects of parasitic capacitances and temperature, and allows the compensating resistors (NIC model) to work over a wider voltage range. Second, we need to balance the loss ($R$) and the gain ($-R$) carefully over a parameter range of $\gamma$ that covers both EPs. Before starting the experiment, our $LC$ circuits are matched to within 0.5\% with an RLC meter. Then, we follow the adjustment-algorithm from Ref.~\cite{Schindler13} to tune our system in every measurement. 

In the $\mathcal{PT}$-symmetric phase ($\gamma<\gamma_{\mathcal{PT}}$) with well matched initial parameters, we get $\sim 10^4$ stable oscillations with two frequency components $\pm\omega_{1,2}$. By fitting the data for the two voltages $V_{1,2}(t)$,  we extract the real and imaginary parts of the two frequencies $\omega_{1,2}$. Fig~\ref{Phase1}a shows $\rm Re\omega_{1,2}/\omega_0$ as a function of the dimensionless gain-loss strength $\gamma$. We see that as $\gamma$ is increased, the measured frequencies $\rm Re\omega_1$ (blue) and $\rm Re\omega_2$ (red) approach each other and become degenerate at $\gamma_\mathcal{PT}=0.79$, while $\rm Im\omega_{1,2}$ remain at zero, Fig.~\ref{Phase1}b. 

For $\gamma_\mathcal{PT}<\gamma\leq\gamma_0$, complex frequencies $\omega_{1,2}$ are conjugates of each other, and their measured values match the theory (black solid lines) well. In this region, because two amplifying modes $\omega_1$ and $\omega_4=-\omega_1^*$ are present, the measured, oscillating voltages $V_{1,2}(t)$ increase exponentially with time before gain saturation. Due to the presence of the two, equally amplifying modes, the contribution from the remaining two, decaying modes $\omega_2=\omega_1^*$ and $\omega_3=-\omega_2^*$ cannot be extracted from the data. After the second EP at $\gamma_0=2.37$, only the contribution of the fastest-amplifying mode can be extracted as it dominates the exponential-growth response before saturation. Fig~\ref{Phase1}b shows that the measured $\rm Im\omega_1$ values match the theory (black solid lines) well, and Fig.~\ref{Phase1}a shows that all the eigenvalues have a zero real part. Thus, our $\mathcal{PT}$-symmetric electric circuit can be experimentally, controllably probed deep into the $\mathcal{PT}$-symmetry broken phase with $\gamma\geq\gamma_0\approx 3\gamma_\mathcal{PT}$.

\section{Discussion}
\label{sec:disc}
In this paper, we have presented theoretical analysis and experimental observation of $\mathcal{PT}$ transitions in a minimal electric circuit with balanced gain and loss. In addition presenting to a transparent way to map the Kirchoff law equations into a Schrödinger equation for the circuit energy dynamics, we have emphasized the four-mode nature of the $\mathcal{PT}$-symmetric electric circuit which gives rise to three regions separated by two exceptional points. We have then presented experimental data for the complex eigenfrequencies of this model, that are in good agreement with the theoretical model. 

Generally speaking, the balanced gain and loss setup in $\mathcal{PT}$-symmetric electric circuit has many advantages over the passive $\mathcal{PT}$-symmetric electric circuit~\cite{Montiel18}, particularly in the large $\gamma$ regime. The latter has no exceptional points, and due to the noise floor for the voltage (and current) data acquisition, one cannot obtain data for sufficiently long time. This makes accurately probing the emergence of slowly decaying eigenmodes in the absence of EPs~\cite{Montiel18,Harter18} truly challenging. With the help of the gain channel, this limitation is overcome.   


\section*{Acknowledgements}
We thank Zhenhua Yu for discussions. JLi received supports from National Natural Science Foundation of
China (NSFC) under Grant No. 11804406, Fundamental Research Funds for Sun Yat-sen University, Science and Technology Program of Guangzhou 2019-030105-3001-0035. LL received supports from NSFC under Grant No. 11774436, Guangdong Province Youth Talent Program under Grant No.2017GC010656, Sun Yat-sen University Core Technology Development Fund, and the Key-Area Research and Development Program of GuangDong Province under Grant No. 2019B030330001. 

\section*{Author contribution}
T. Wang and J. Fang contributed equally to this work.  Correspondence and requests for materials should be addressed to Y.N.J., J.L., or L.L. 


%
%


\end{document}